\documentclass[twoside,twocolumn,9pt]{article}
\usepackage{extsizes}
\usepackage[super,sort&compress,comma]{natbib}
\usepackage[version=3]{mhchem}
\usepackage[left=1.5cm, right=1.5cm, top=1.785cm, bottom=2.0cm]{geometry}
\usepackage{balance}
\usepackage{widetext}
\usepackage{times,mathptmx}
\usepackage{sectsty}
\usepackage{graphicx}
\usepackage{lastpage}
\usepackage[format=plain,justification=raggedright,singlelinecheck=false,font={stretch=1.125,small,sf},labelfont=bf,labelsep=space]{caption}
\usepackage{float}
\usepackage{fancyhdr}
\usepackage{fnpos}
\usepackage[english]{babel}
\usepackage{array}
\usepackage{droidsans}
\usepackage{charter}
\usepackage[T1]{fontenc}
\usepackage[usenames,dvipsnames]{xcolor}
\usepackage{setspace}
\usepackage[compact]{titlesec}
\usepackage{srctex}
\usepackage{hyperref}

\definecolor{cream}{RGB}{222,217,201}

\begin{document}

\pagestyle{fancy}
\thispagestyle{plain}
\fancypagestyle{plain}{

\fancyhead[C]{\includegraphics[width=18.5cm]{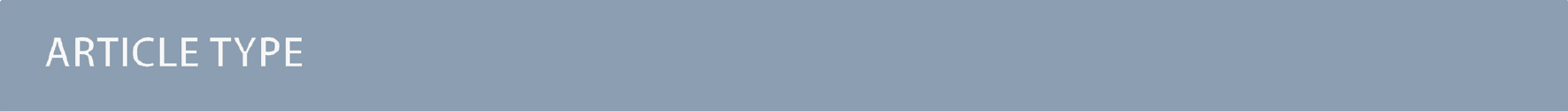}}
\fancyhead[L]{\hspace{0cm}\vspace{1.5cm}\includegraphics[height=30pt]{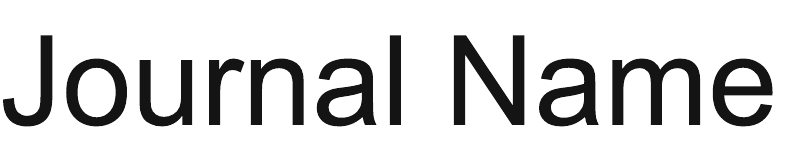}}
\fancyhead[R]{\hspace{0cm}\vspace{1.7cm}\includegraphics[height=55pt]{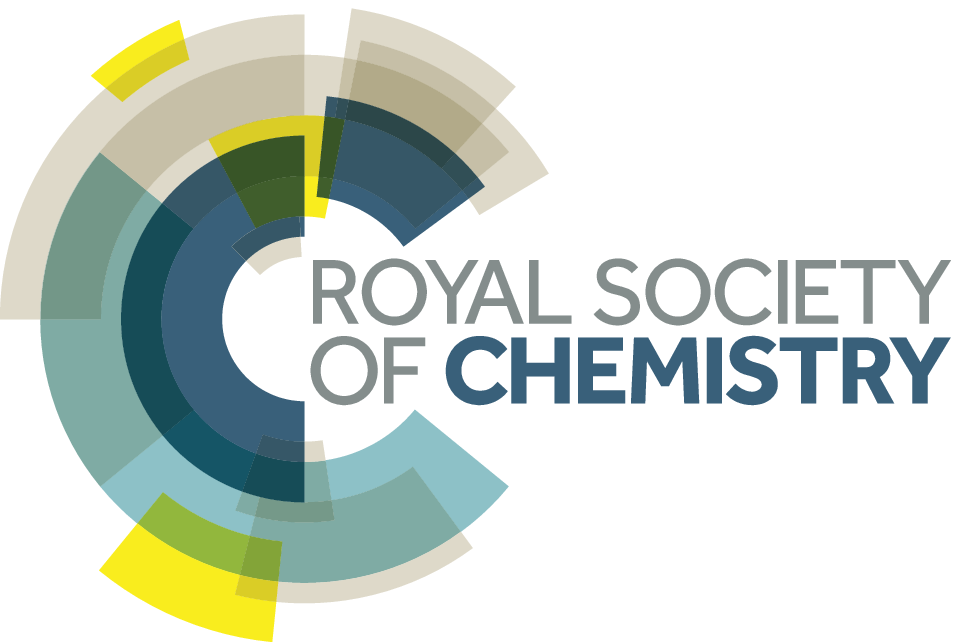}}
\renewcommand{\headrulewidth}{0pt}
}

\makeFNbottom
\makeatletter
\renewcommand\LARGE{\@setfontsize\LARGE{15pt}{17}}
\renewcommand\Large{\@setfontsize\Large{12pt}{14}}
\renewcommand\large{\@setfontsize\large{10pt}{12}}
\renewcommand\footnotesize{\@setfontsize\footnotesize{7pt}{10}}
\makeatother

\renewcommand{\thefootnote}{\fnsymbol{footnote}}
\renewcommand\footnoterule{\vspace*{1pt}%
\color{cream}\hrule width 3.5in height 0.4pt \color{black}\vspace*{5pt}}
\setcounter{secnumdepth}{5}

\makeatletter
\renewcommand\@biblabel[1]{#1}
\renewcommand\@makefntext[1]%
{\noindent\makebox[0pt][r]{\@thefnmark\,}#1}
\makeatother
\renewcommand{\figurename}{\small{Fig.}~}
\sectionfont{\sffamily\Large}
\subsectionfont{\normalsize}
\subsubsectionfont{\bf}
\setstretch{1.125} 
\setlength{\skip\footins}{0.8cm}
\setlength{\footnotesep}{0.25cm}
\setlength{\jot}{10pt}
\titlespacing*{\section}{0pt}{4pt}{4pt}
\titlespacing*{\subsection}{0pt}{15pt}{1pt}

\fancyfoot{}
\fancyfoot[LO,RE]{\vspace{-7.1pt}\includegraphics[height=9pt]{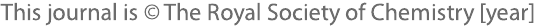}}
\fancyfoot[CO]{\vspace{-7.1pt}\hspace{13.2cm}\includegraphics{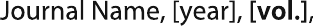}}
\fancyfoot[CE]{\vspace{-7.2pt}\hspace{-14.2cm}\includegraphics{RF}}
\fancyfoot[RO]{\footnotesize{\sffamily{1--\pageref{LastPage} ~\textbar  \hspace{2pt}\thepage}}}
\fancyfoot[LE]{\footnotesize{\sffamily{\thepage~\textbar\hspace{3.45cm} 1--\pageref{LastPage}}}}
\fancyhead{}
\renewcommand{\headrulewidth}{0pt}
\renewcommand{\footrulewidth}{0pt}
\setlength{\arrayrulewidth}{1pt}
\setlength{\columnsep}{6.5mm}
\setlength\bibsep{1pt}

\makeatletter
\newlength{\figrulesep}
\setlength{\figrulesep}{0.5\textfloatsep}

\newcommand{\topfigrule}{\vspace*{-1pt}%
\noindent{\color{cream}\rule[-\figrulesep]{\columnwidth}{1.5pt}} }

\newcommand{\botfigrule}{\vspace*{-2pt}%
\noindent{\color{cream}\rule[\figrulesep]{\columnwidth}{1.5pt}} }

\newcommand{\dblfigrule}{\vspace*{-1pt}%
\noindent{\color{cream}\rule[-\figrulesep]{\textwidth}{1.5pt}} }

\makeatother

\twocolumn[
  \begin{@twocolumnfalse}
\vspace{3cm}
\sffamily
\begin{tabular}{m{4.5cm} p{13.5cm} }

\includegraphics{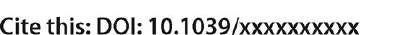} & \noindent\LARGE{\textbf{Universal self-assembly of one-component three-dimensional dodecagonal quasicrystals}} \\
\vspace{0.3cm} & \vspace{0.3cm} \\

 & \noindent\large{Roman Ryltsev,$^{\ast}$\textit{$^{ab}$} Nikolay Chtchelkatchev\textit{$^{cde}$}} \\

\includegraphics{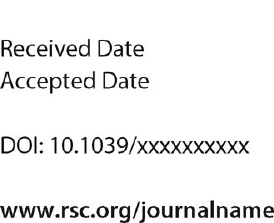} & \noindent\normalsize{Using molecular dynamics simulations, we study computational self-assembly of one-component three-dimensional dodecagonal (12-fold) quasicrystals in systems with two-length-scale potentials.  Existing criteria for three-dimensional quasicrystal formation are quite complicated and rather inconvenient for particle simulations. So to localize numerically the quasicrystal phase, one should usually simulate over a wide range of system parameters. We show how to universally localize the parameters values at which dodecagonal quasicrystal order may appear for a given particle system. For that purpose, we use a criterion recently proposed for predicting decagonal quasicrystal formation in one-component two-length-scale systems. The criterion is based on two dimensionless effective parameters describing the fluid structure which are extracted from radial distribution function. The proposed method allows reducing the time spent for searching the parameters favoring certain solid structure for a given system. We show that the method works well for dodecagonal quasicrystals; this results is verified on four systems with different potentials: Dzugutov potential, oscillating potential which mimics metal interactions, repulsive shoulder potential describing effective interaction for core/shell model of colloids and embedded-atom model potential for aluminum. Our results suggest that mechanism of dodecagonal quasicrystal  formation is universal for both metallic and soft-matter systems and it is based on competition between interparticle scales.} \\

\end{tabular}

 \end{@twocolumnfalse} \vspace{0.6cm}

  ]

\renewcommand*\rmdefault{bch}\normalfont\upshape
\rmfamily
\section*{}
\vspace{-1cm}

\footnotetext{\textit{$^{a}$~Institute of Metallurgy, UB RAS, 620016, 101 Amundsen str., Ekaterinburg, Russia. E-mail: rrylcev@mail.ru}}
\footnotetext{\textit{$^{b}$~Ural Federal University, 620002, 19 Mira str,, Ekaterinburg, Russia.}}
\footnotetext{\textit{$^{c}$~L.D. Landau Institute for Theoretical Physics, RAS, 142432, Ac. Semenov 1-A, Chernogolovka, Russia.}}
\footnotetext{\textit{$^{d}$~All-Russia Research Institute of Automatics, 22 Suschevskaya, Moscow 127055, Russia}}
\footnotetext{\textit{$^{e}$~Moscow Institute of Physics and Technology,141700, Institutskiy per.9, Dolgoprudny, Moscow Region, Russia}}



\section{Introduction}

Quasicrystals (QCs) have been experimentally observed for both  metallic alloys~\cite{stadnik2012QC,Tsai2008SciTechAdvMat} and soft matter systems~\cite{Zeng2004Nature,Fischer2011PNAS,Hayashida2007PRL,Talapin2009Nature,Zaidouny2014SoftMatt,Ungar2005SoftMatt} that suggests a common microscopic mechanism of QC formation.

 Stability of three-dimensional (3D) one-component QCs has been theoretically predicted using density functional theory ~\cite{Denton1998PRL,Denton1997PRB} and then confirmed by molecular dynamics simulations~\cite{Dzugutov1993PRL,Englel2015Nature,Ryltsev2015SoftMatt,Damasceno2017JPCM,Metere2016SoftMatter}. The results obtained in these papers suggest that general idea explaining QC formation is the existence of two ore more interparticle length-scales. This idea is supported by the fact that effective interactions for metallic ~\cite{Lee1981Book,Mitra1978JPhysC,Mihalkovich2012PRB,Dubinin2014RusChemRev,Dubinin2014JNonCrystSol} and soft matter systems~\cite{Likos2001PhysRep,Watzlawek1999PRL,Likos2002JPhysCondMatt,Prestipino2009SoftMatt,Rechtsman2006PRE} are often described by multi-length-scale potentials.

 A general problem of computer simulation of 3D QCs is the lack of simple geometrical criteria of QC formation. So to localize numerically the QC phase, one should usually simulate over a wide range of system parameters \cite{Englel2015Nature,Damasceno2017JPCM}.  Similar problem exists for complex crystal phases in systems with multi-scale interactions~\cite{Englel2015Nature}.  Thus, an universal procedure allowing to predict somehow the formation of complex solid structures (including QCs) is extremely urgent.

   Recently, we have proposed a method to predict self-assembly of decagonal QCs in one-component two-length-scale systems \cite{Ryltsev2015SoftMatt}. The method suggests the formation of QCs from the fluid phase is mostly determined by the values of two dimensionless structural parameters of the fluid.  The parameters reflect the existence of two effective interparticle distances (bond lengthes) originated from two-length-scale nature of interaction potential. These are the ratio between effective bond lengthes, $\lambda$, and the fraction of short-bonded particles $\phi$. It has been shown that the criterion proposed is robust under change of potential and may be applicable to any system with two-length-scale interaction.

Here we show that the criterion works well for the case of dodecagonal (12-fold) quasicrystals (DQCs). In order to show that, we use four different two-length-scale potentials: Dzugutov potential \cite{Dzugutov1993PRL} and oscillating pair potential (OPP) \cite{Englel2015Nature}  which mimic oscillating metal interactions, repulsive shoulder system (RSS) potential \cite{Ryltsev2013PRL} corresponding to core/shell model of colloids and the embedded-atom model (EAM) potential for aluminum proposed in \cite{Mishin1999PRB}. The values of effective parameters favoring dodecagonal order are determined from the system with Dzugutov potential for which temperature-density domain of DQC formation is known \cite{Dzugutov1993PRL}. Adjusting the states of RSS and OPP fluids to obtain the same values of effective parameters, we observe self-assembly of the same DQC phases at cooling. The values of the parameters for EAM model \cite{Mishin1999PRB} of liquid aluminum near the liquid-DQC transition reported in \cite{Prokhoda2014arXiv} are also the same. This result suggests common nature of both metallic and soft matter DQCs arising from competition between length scales.

\begin{figure}
  \centering
  \includegraphics[width=0.7\columnwidth]{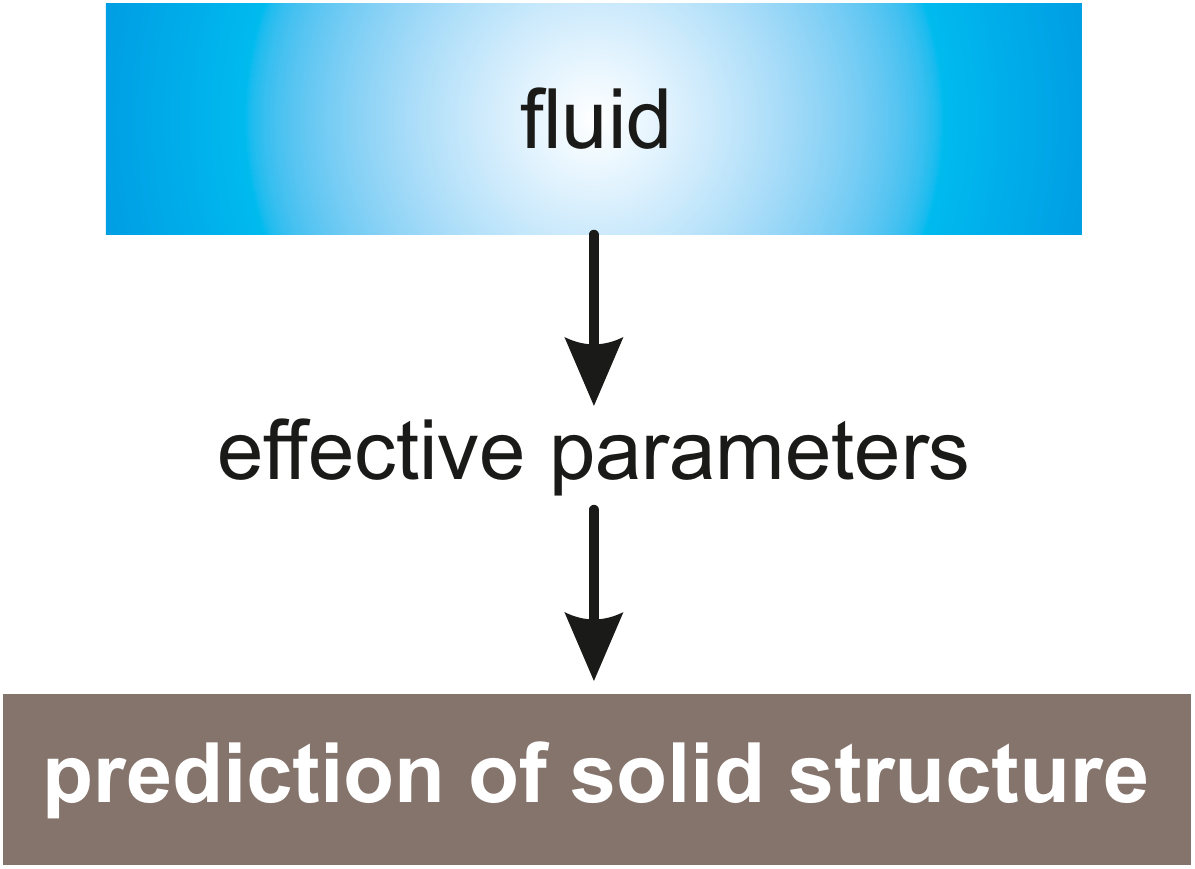}\\
 \caption{General scheme to obtain solid state structure of two-length-scale systems with using effective parameters}
  \label{fig:scheme}
\end{figure}

The proposed method allows reducing the time spent for searching the parameters favoring certain solid structure for two-length scale systems. Given the value of effective parameters favoring the formation of some structure, we can predict if this structure self-assemblies from the fluid at cooling (see the general scheme in Fig.~\ref{fig:scheme} for explanation).

\section{Methods\label{sec:methods}}

\subsection{Interparticle potentials}

We investigate by the molecular dynamics simulations one-component 3D systems of particles interacting via four different two-length-scale potentials (see Fig.~\ref{fig:rdf_pot}a). The first one is well-known Dzugutov potential\cite{Dzugutov1992PRA}:

\begin{equation}\label{Dzug_pot}
U_{\rm dz}=U_1(r)+U_2(r),
\end{equation}
where
\begin{equation}\label{Dzug_U1}
U_1 (r) = \left\{ \begin{array}{l}
 A(r^{ - m}  - B)\exp (c/(r - a)),\quad r < a \\
 0,\quad r \ge a, \\
 \end{array} \right.
\end{equation}
and
\begin{equation}\label{Dzug_U1}
U_2 (r) = \left\{ \begin{array}{l}
 B\exp [d/(r - b)],\quad r < b \\
 0,\quad r \ge b, \\
 \end{array} \right.
\end{equation}
with the parameters $m=16$, $A=5.82$, $c=1.1$, $a=1.87$, $B=1.28$, $d=0.27$, $b=1.94$. A one-component system of particles interacting with the potential \ref{Dzug_pot}--\ref{Dzug_U1} can form DQC phase \cite{Dzugutov1993PRL} as well as QC approximants \cite{Keys2007PRL, Denton2000PRE} and non-trivial crystal structures \cite{Denton2000PRE}.

The second potential we use is the repulsive shoulder system (RSS) potential:~\cite{Ryltsev2013PRL}
\begin{equation}\label{rss}
U_{\rm rss}(r)=\varepsilon\left({d}/{r}\right)^{n}+\varepsilon {\rm n_f}\left[2k_0\left(r-\sigma \right)\right],
\end{equation}
where ${\rm n_f}(x)=1/[1+\exp(x)]$, $\varepsilon$ is the unit of energy, $d$ and $\sigma$ are ``hard''-core and ``soft''-core diameters.

The hard-core analog of RSS potential was developed by Adler and Yong to explain melting curve extrema \cite{Young1977PRL}. Later it was rediscovered by Stishov~\cite{Stishov2002PhilMag} and applied in smooth form, eq.\ref{rss}, to describe phase diagrams with polymorphous transitions, water-like anomalies, glassy dynamics and formation of decagonal QC, see Refs.~\cite{Ryltsev2013PRL, Ryltsev2015SoftMatt} and references therein. He we take $n=14$, $k_0=10$, and $\sigma=1.75$ to produce the same values of effective parameters as those for Dzugutov potential (see sec. \ref{sec:results}).

The third potential used is the modified oscillating pair potential (OPPm):
\begin{equation}
U_{\rm opp}(r)=1/r^{15}+a\exp(-(r/b)^m)\cos(kr-\varphi)
\end{equation}
with $a=0.5$, $b=1.45$, $m=20$, $k=14.4$, $\varphi=17.125$. It is slightly modified potential which was first introduced in \cite{Mihalkovich2012PRB} and then used to simulate icosahedral QCs~\cite{Englel2015Nature}. In contrast to original OPP from Ref.~\cite{Englel2015Nature},  we have just replaced pre-cosine power factor by exponential one to suppress oscillations after second minimum (to restrict the system by only two characteristic length scales). The values of parameters $a$, $b$, $m$, $k$ reported above have been chosen to provide long/short bond length ratio $\lambda\sim 1.7$ that is optimal for DQC formation (see sec. \ref{sec:results}).

Finally, the forth potential used is the EAM one proposed in Ref.~\cite{Mishin1999PRB} for aluminum. Within the frameworks of EAM, the potential energy of the system $E_{\rm pot}$ is represented as the sum of the pair interaction contribution $U_{\rm pair}$ and the embedding energy $F(\rho)$ depending on local electron density $\rho$. So a EAM potential is effectively a many-body one. To compare visually the aluminum EAM potential with pair two-length-scale potentials described above, we use effective pair format for EAM $U_{\rm eff}=U_{\rm pair}(r)+F(\rho(r))$ taking into account the distance dependence of electron density \cite{Mishin1999PRB}. Note that simulations were performed with the original many-body formulation of aluminum EAM.

 \begin{figure*}
  \centering
  \includegraphics[width=0.7\textwidth]{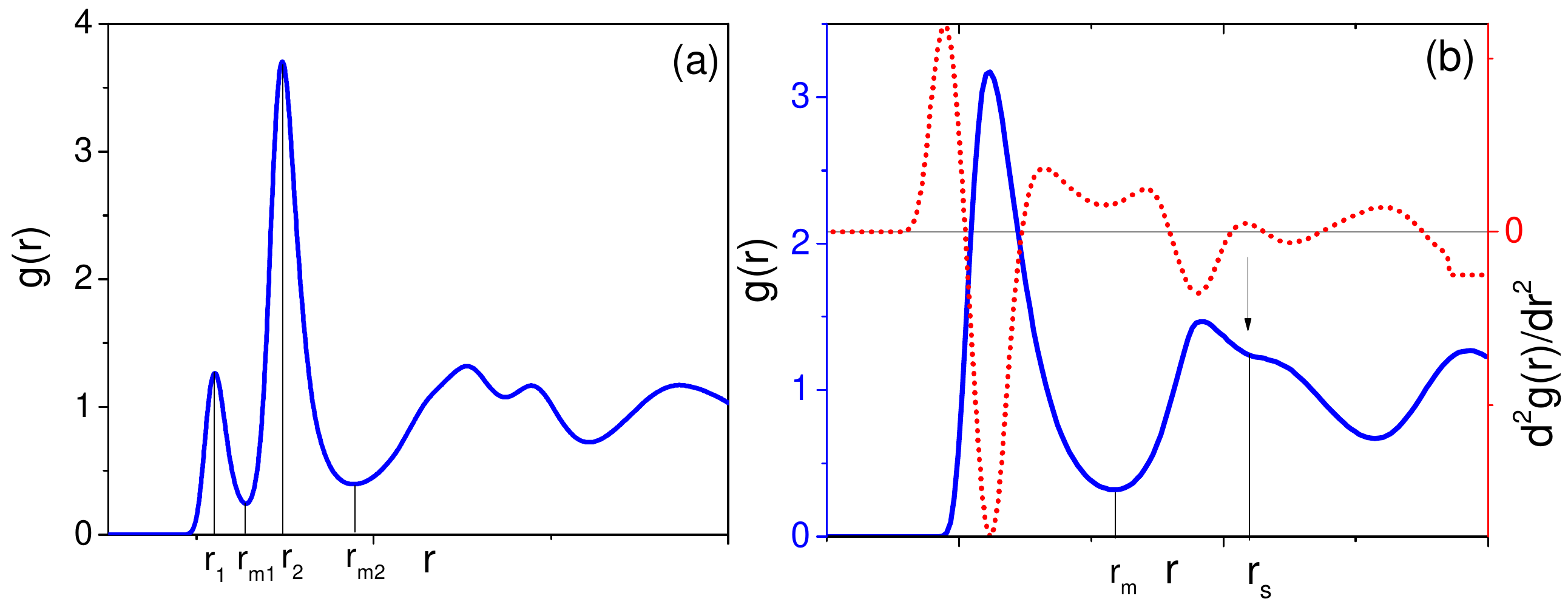}\includegraphics[width=0.27\textwidth]{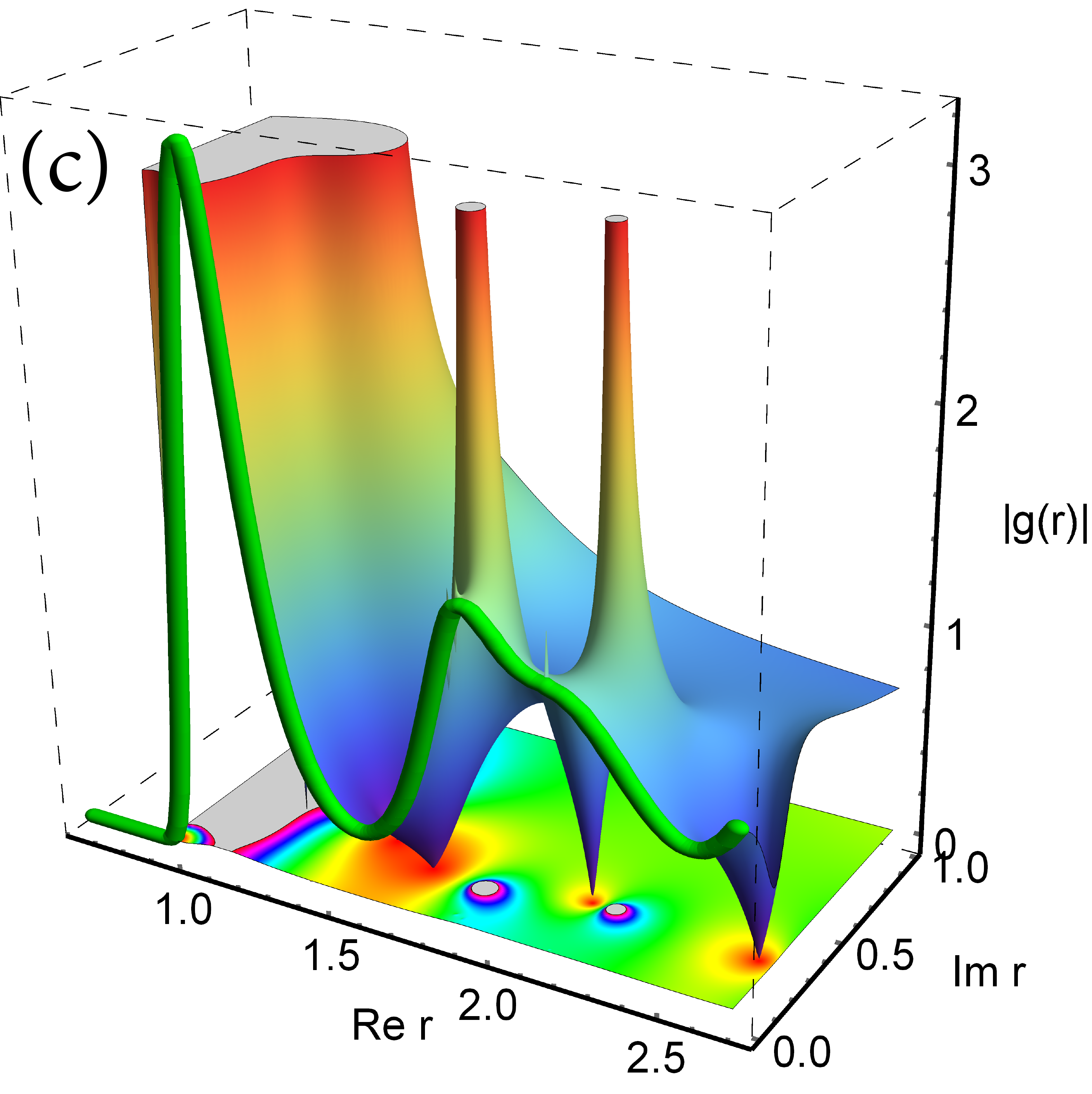}\\
 \caption{Fluid state radial distribution functions of different two-length-scale systems demonstrating the definition of effective parameters. (a) RSS with the parameters corresponding to decagonal QC: $\sigma=1.37$, $\rho=0.474$, $T=0.11$. We see excellent separation of $g(r)$ peaks and so effective parameters are well defined.  (b) System with Dzugutov potential with $\rho=0.85$, $T=0.6$. In this case, the splitting of the second $g(r)$ peak means the overlapping of the second coordination shell of short-bonded particles and the first coordination shell of the long-bonded ones. To separate the second peak, the second derivative of $g(r)$ may be used; its maximum allows estimating the distance corresponding to intersection of the subpeaks. (c) The three dimensional graph with the density-plot projection of analytical continuation of real-argument $g(r)$ shown as the green ``tube''. The figure illustrates another, unambiguous, way to extract subpeaks from the second $g(r)$ peak: we analytically continued $g(r)$ into complex plain $(\mathrm{Re}\, r,\mathrm{Im}\, r)$, taking $r$ as complex variable of the complex-valued function that is equal $g(r)$ at real $r$. Real and imaginary coordinates of the poles (peaks in 3D graph) in complex plain of $r$ give the centers of the $g(r)$ subpeaks in (b) and the subpeak width correspondingly.}
  \label{fig:eff_par}
\end{figure*}

\subsection{Simulation details}
  Hereafter we use dimensionless units like Lennard-Jones ones that is normalizing the energy, temperature and distance by the corresponding potential parameters. For example, for RSS we have $\tilde{{\bf r}}\equiv {\bf r}/d$, $\tilde U=U/\varepsilon$, temperature $\tilde T=T/\epsilon$, density $\tilde{\rho}\equiv N d^{3}/V$, and time $\tilde t=t/[d\sqrt{m/\varepsilon}]$, where $m$ and $V$ is the molecular mass and system volume correspondingly. For the EAM model of aluminum, the value of effective pair potential $U_{\rm eff}$ at the first minimum (see Fig.~\ref{fig:rdf_pot}a) was chosen as the energy unit.

For molecular dynamics simulations, we use $\rm{LAMMPS}$ package~\cite{Plimpton1995JCompPhys,lammps}. The system of $N=20000$ particles was simulated under periodic boundary conditions in Nose-Hoover NVT ensemble. This amount of particles is enough to obtain satisfactory diffraction patterns to study (quasi)crystal symmetry (see Fig.~\ref{fig:snaphsots}). Taking larger system requires too much calculation time necessary to QC equilibration. The molecular dynamics time step was $\delta t=0.003-0.01$ depending on system temperature \cite{Kuksin2005MolSim, Norman2001JETP}.

To study solid phases, we cooled the system starting from a fluid in a stepwise manner and completely equilibrated at each step. The time dependencies of temperature, pressure and configurational energy were analyzed to control equilibration \cite{Kuksin2005MolSim}.

To study the structure of both fluid and solid phases we use radial distribution functions $g(r)$, bond order parameters $q_l$~\cite{Steinhardt1981PRL,Steinhardt1983PRB,Hirata2013Science}, diffraction analysis and visual analysis of the snapshots. Detailed description of these methods as well as the procedure for preparing and relaxing the solid phases is presented in Ref.~\cite{Ryltsev2015SoftMatt}

\section{Results and discussion\label{sec:results}}

\subsection{Effective parameters}

Earlier, we have proposed \cite{Ryltsev2015SoftMatt} that the structure of low-temperature solid phases in one-component two-length-scale systems is essentially determined by two dimensionless parameters of the high-temperature fluid state. The parameters are the ratio between effective interparticle distances (bond lengthes), $\lambda$, and the fraction of short-bonded particles, $\phi$. These parameters can be extracted from radial distribution function $g(r)$  of a fluid. Indeed, the existence of two length scales causes splitting of the first peak in  $g(r)$  (see Fig.~\ref{fig:eff_par}a). So we have $\lambda=r_2/r_1$, where $r_1$ and $r_2$ are the positions of the $g(r)$ subpeak maxima. The bond fraction is determined as $\phi=n_1/(n_1+n_2)$ , where $n_1  = 4\pi\rho \int_0^{r_{m1}} {r^2 g(r)dr}$ and $n _2  = 4\pi\rho \int_{r_{m1}}^{r_{m2}} {r^2 g(r)dr}$ are respectively the effective numbers of short- and long-bonded particles in the first coordination shell. Here, $r_{m1}$ and $r_{m2}$ are locations of the first and the second $g(r)$ minima separating the subpeaks (Fig.~\ref{fig:eff_par}a).

The effective parameters are well defined in the case of $1.2 <\lambda < 1.6$ then $g(r)$ subpeaks corresponding to short- and long-bonded particles are perfectly separated at arbitrary $\phi$ values. For example, in Fig.~\ref{fig:eff_par}a, we show $g(r)$ of RSS fluid for the effective parameters $\lambda=1.37$, $\phi=0.474$ corresponding to decagonal QC\cite{Ryltsev2015SoftMatt}. But the situation is more complicated in the case of DQC considering here. Indeed, in Fig.~\ref{fig:eff_par}b  we show $g(r)$ for the system with Dzugutov potential with the parameters corresponding to fluid slightly above the fluid-DQC transition \cite{Dzugutov1993PRL}. As seen from the picture, the value of $\lambda$ is about 1.7 that means the first coordination shell of long-bonded particles overlaps with the second coordination shell of short-bonded ones (see splitting of the second $g(r)$ peak in Fig.~\ref{fig:eff_par}b). To determine effective parameters in this case, we use the method of peak separation widely used in spectroscopy \cite{Butler1980PhotocemPotobiol,Arag2008JBrazChemSoc}. The method is based on using high order (2th and 4th) derivatives to separate overlapped peaks. In Fig.~\ref{fig:eff_par}b we show the second derivative of $g(r)$ for the Dzugutov potential. As seen from the picture, the maximum of $d^2 g(r)/dr^2$ allows estimating the distance $r_s$ corresponding to intersection of the subpeaks.  So the effective numbers of short- and long-bonded particles can be estimated as $n_1  = 4\pi\rho \int_0^{r_{m}} {r^2 g(r)dr}$ and $n _2  = 4\pi\rho \int_{r_{m}}^{r_{s}} {r^2 g(r)dr}$.

Figure.~\ref{fig:eff_par}b can cause a feeling that the separation of the second $g(r)$ peak into two subpeaks is an artificial, non-physical and mathematically fragile procedure. We describe below a method that makes one sure that the second $g(r)$ peak indeed has two subparts. We consider $g(r)$ as the reduction of some complex-valued function of complex variable $r$ on real  axes.  Using  analytical  continuation procedure, we can reconstruct this complex-valued function using $g(r)$ as the ``source''. We perform analytical continuation of $g(r)$  into complex $(\mathrm{Re}\, r,\mathrm{Im}\, r)$ plain numerically using Pade-approximants, see Refs.~\cite{baker1996pade,Schott2016PhysRevB,Chtchelkatchev2015JETPLett} for the details of the procedure. In Fig.~\ref{fig:eff_par}c, we show the three dimensional graph with the density-plot projection of analytical continuation of $g(r)$ presented in Fig.~\ref{fig:eff_par}b where $g(r)$ at real $r$ is shown as the green ``tube''. We see that, in complex-$r$ plain, the second $g(r)$ peak transforms into two peaks. Detailed investigation shows that these peaks correspond to poles where the complex-valued function diverges like $1/(r-R_i)$, where $R_i$, $i=1,2$ are the complex coordinates of the poles. If we return to real-number ``physical'' coordinates then $1/(r-R_i)$-contribution transforms into the Lorentzian peak: $1/[(r-\mathrm{Re}\,R_i)^2+(\mathrm{Im}\,R_i)^2]$, so $\mathrm{Re}\,R_i$ and $\mathrm{Im}\,R_i $ become the center and the width of the Lorentzian, respectively~\cite{Chtchelkatchev2015JETPLett}. Two poles like shown in Fig.~\ref{fig:eff_par}c transform into a superposition of two Lorentzian peaks at real values of $r$. Thus the second peak of $g(r)$ in Fig.~\ref{fig:eff_par}b is indeed the superposition of two distinct subpeaks.

The very fact that the pair correlation function reduces in the main approximation to the superposition of Lorentzians functions rather than, for example, the gaussian functions, is quite interesting observation that can be generalized to many other systems \cite{Chtchelkatchev2016JETPLett,Khusnutdinoff2016JETP}. This issue and technical details will be described in a separate paper.

Using the methods described above, the effective parameters of the system with Dzugutov potential slightly above the fluid-DQC transition have been estimated to be $\lambda=1.74$, $\phi=0.42$. These values will be further used as reference ones to obtain DQCs in other two-length-scale systems under consideration.

\begin{figure*}
  \centering
  \includegraphics[width=\textwidth]{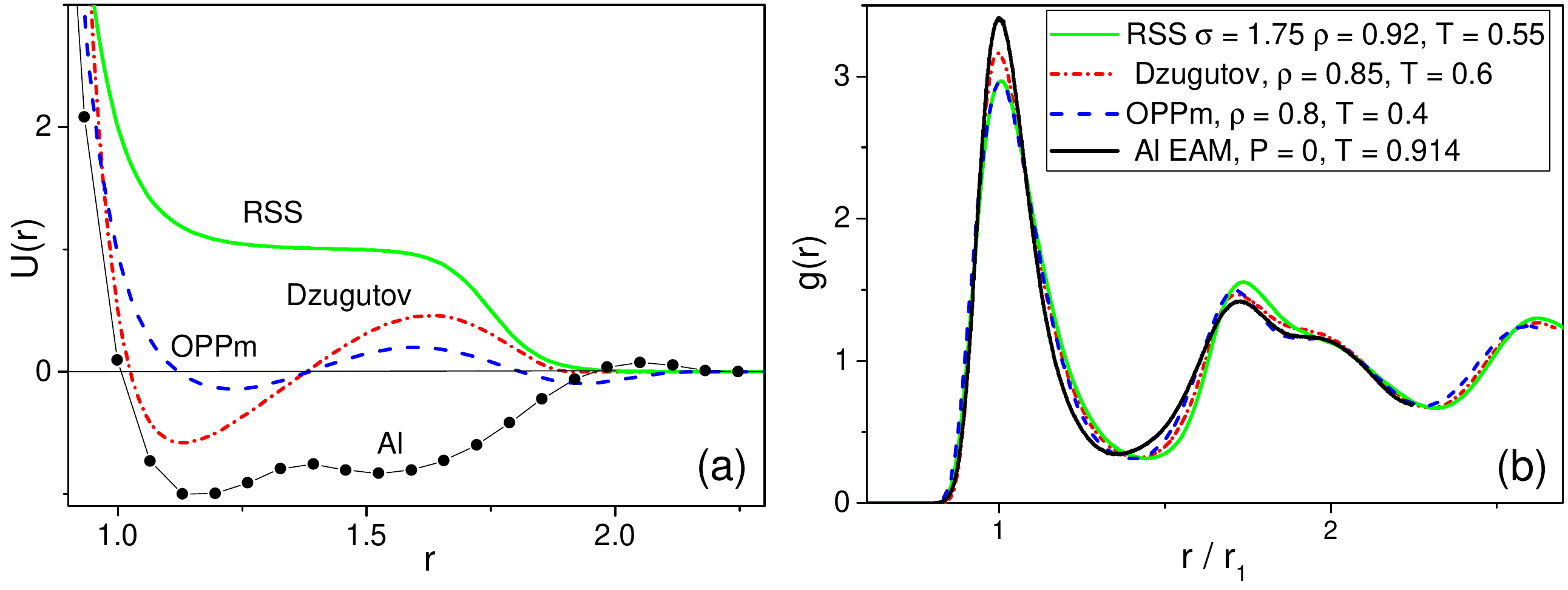}\\
 \caption{(a) Pair potentials of different two-length-scale systems demonstrating DQC formation. For the aluminum, the effective pair potential, constructed from EAM one, is shown. (b) Fluid state radial distribution functions of the systems with similar values of effective parameters favoring self-assembly of DQC.}
  \label{fig:rdf_pot}
\end{figure*}

\subsection{Universal self-assembly of DQC}

\begin{figure*}
  \centering
  \includegraphics[width=0.9\textwidth]{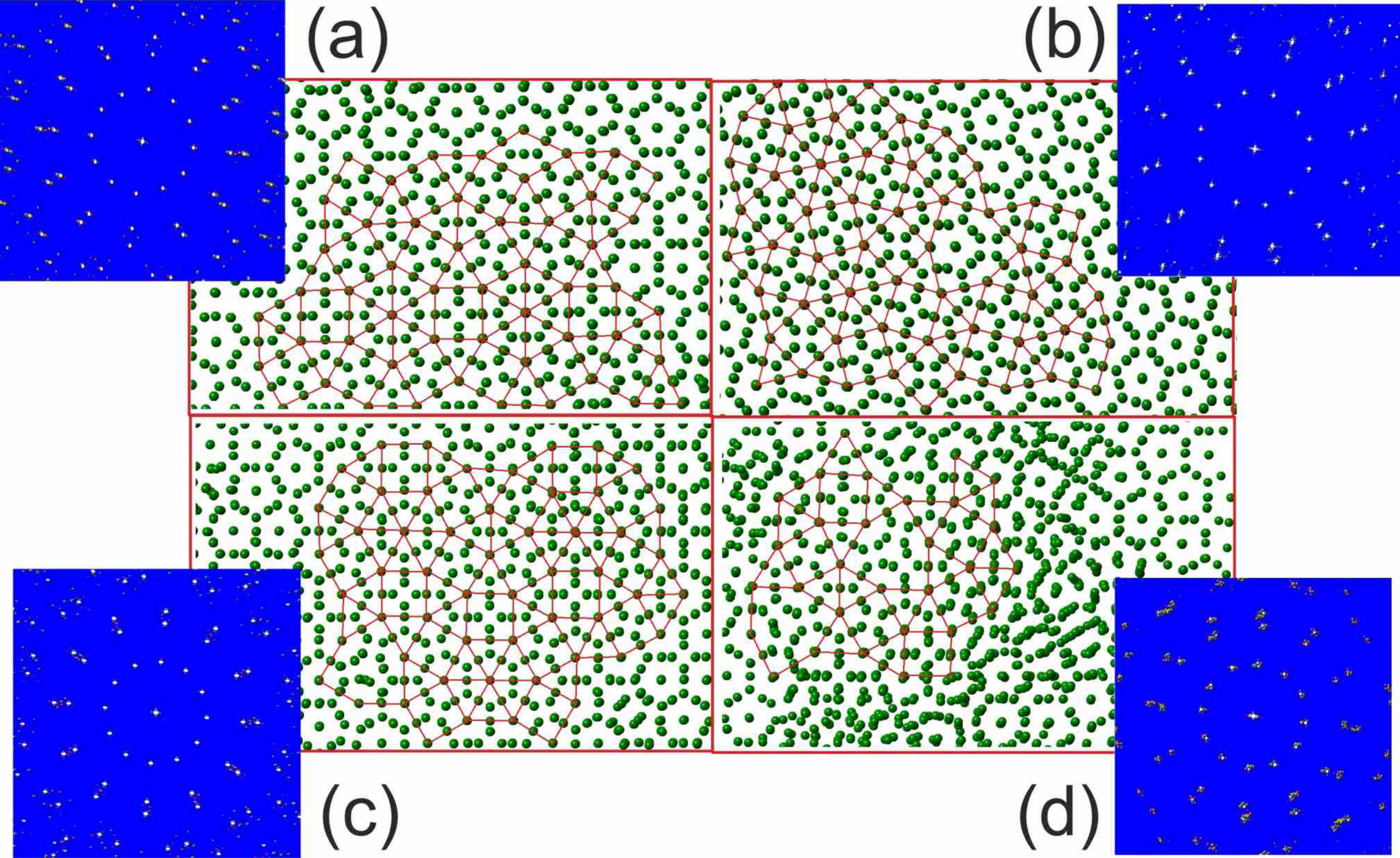}\\
 \caption{Typical atomic configurations of two-scale systems investigated demonstrating dodecagonal order (a) System with Dzugutov potential with $\rho=0.85$, $T=0.55$; (b) RSS with  $\sigma=1.75$, $\rho=0.92$, $T=0.53$; (c) OPPm system with $\rho=0.8$, $T=0.35$; (d) EAM potential for Al at $P=0$, $T=0.914$ ($T=700$ K). Red lines connecting the centers of dodecagons demonstrate QC-like tiling.  Insets in the corners show corresponding diffraction patterns with 12-fold symmetry.}
  \label{fig:snaphsots}
\end{figure*}

To validate the universality of effective parameters values estimated for DQC formation from Dzugutov system, we tune the parameters of both RSS and OPPm systems to obtain similar RDF in fluid phase (see Fig.~\ref{fig:rdf_pot}) and so the similar values of effective parameters. The value of the ratio between short and long bond lengths $\lambda$ can be tuned by varying either the core/shall ratio $\sigma$ in the case of RSS or the distance between potential minima for OPPm. The value of short bond concentration $\phi$ mostly depends on system density.
  We also calculated the effective parameters for the aluminum with EAM potential proposed in \cite{Mishin1999PRB} at the thermodynamic state near the liquid-DQC transition \cite{Prokhoda2014arXiv}. The values of effective parameters for the systems under consideration were obtained to be $(\lambda=1.73, \phi=0.38)$ for RSS; $(\lambda=1.7,\phi=0.42)$ for OPPm and $(\lambda=1.73, \phi=0.36)$  for aluminum. We see that the $(\lambda,\phi)$  values for all the systems are very close to each other.

The systems with the parameters, chosen as described above, were cooled down from the fluid phase till the fluid-solid transition occurs. The resulting solid state in all cases consisted of few highly ordered quasicrystalline grains with pronounced dodecagonal symmetry. Typical snapshots of such DQC grains in the plain orthogonal to 12-fold axis are  presented in Fig.\ref{fig:snaphsots}. We see that all the systems demonstrate the same dodecagonal structure. The diffraction patterns of each structure are also shown to demonstrate the identical 12-fold symmetry of the samples.

Hereafter we use the term DQC having in mind the system may also fall into a crystalline approximant with local QC symmetry. Moreover, any QC-like configuration constrained by periodic
boundary conditions is in fact a periodic approximant in the sense of the global order. It should be also noted that DQC phase observed may be not the thermodynamically stable one for the systems under consideration. For example, it is known that, for the Dzugutov potential system, the DQC phase is thermodynamically metastable with respect to the $\sigma$-phase periodic approximant \cite{Denton2000PRE}. The study of thermodynamic stability of DQC phases observed as well as the investigation of subtle structural features like difference between true QC and approximant phases are out of the framework of this paper.  Anyway, the observed DQC structures are physically stable over the time scale available for simulation and have QC structure on the mesoscale of the simulation box; it is enough for the purposes of this work.

  Thus the values of effective parameters favorable to DQC formation are estimated to be $\lambda \sim 1.7$, $\phi \sim 0.4$. These values are only the estimation; in fact DQC phase can form in certain intervals of effective parameters around estimate of this type~\cite{Ryltsev2015SoftMatt}. The exact determination of these intervals for each system under consideration is the matter of separate work.  It should be also noted that the obtained values of effective parameters can be only used to predict the formation of certain type of DQC presented in Fig.~\ref{fig:snaphsots}. Other types of one-component 3D DQC recently observed in computer simulations \cite{Damasceno2017JPCM,Metere2016SoftMatter} have different structure and so different values of effective parameters. The same holds true for recently reported three dimensional decagonal QC whose structure differs from that for decagonal QC obtained in Ref.~\cite{Ryltsev2015SoftMatt}. The study of applicability of the effective parameters method for these new types of one-component QCs is the matter of separate work.

 Note that the DQC structure obtained for aluminum is of much worse quality than that for other systems investigated. Indeed, in Fig.~\ref{fig:snaphsots}d we see a lot of structure defects disturbing the QC structure. That is because we did not tune neither the parameters of EAM potential nor the thermodynamic state for aluminum. The system with original parameters proposed in Refs.~\cite{Mishin1999PRB,Prokhoda2014arXiv} generates fluid  whose $\phi$ value is slightly less than optimal one obtained from the Dzugutov potential (see also the Fig.~\ref{fig:pmf} in the next section).

\begin{figure}
  \centering
  \includegraphics[width=\columnwidth]{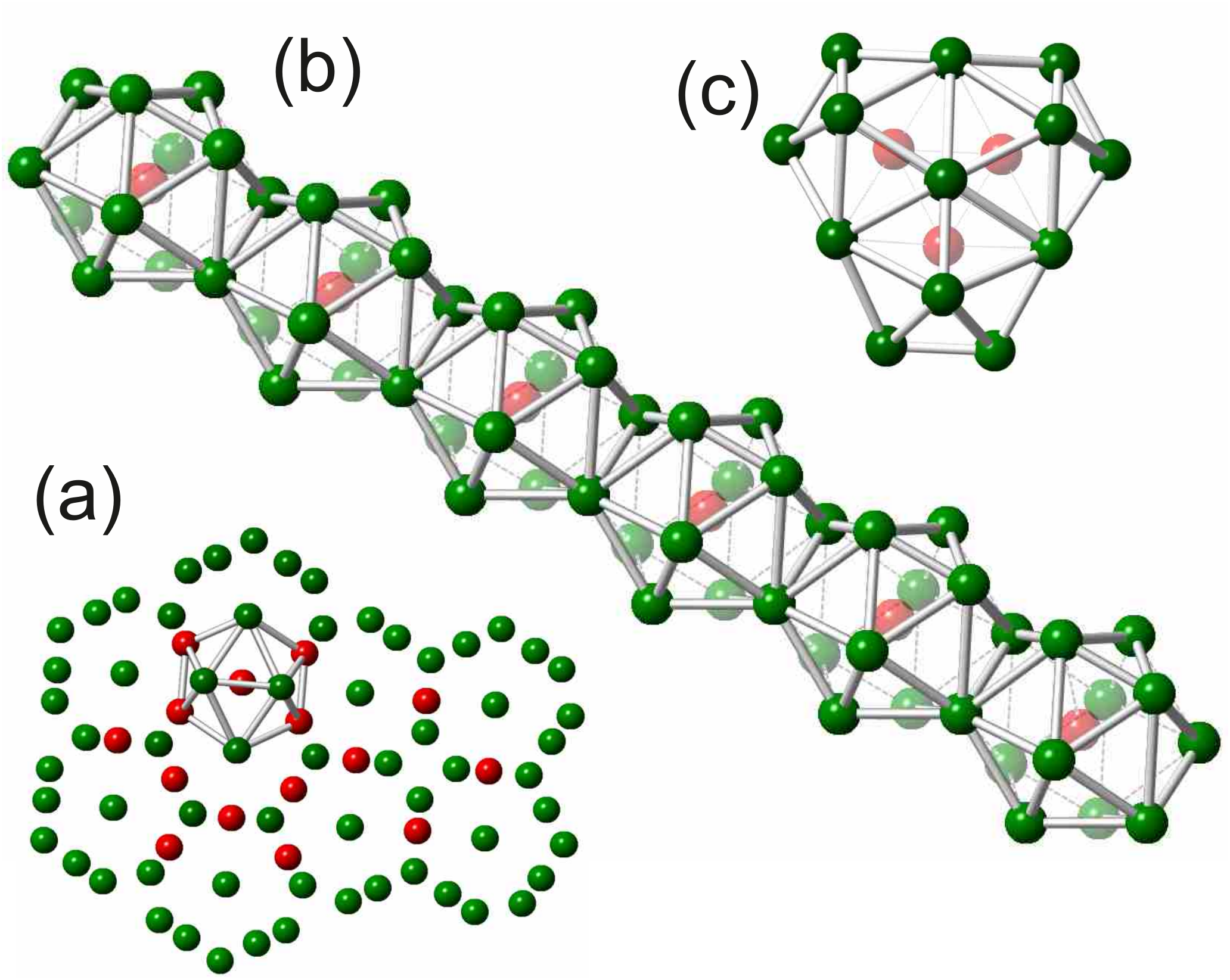}\\
 \caption{ (a) Typical fragment of DQC tiling for RSS system; the particles which are the centers of icosahedra are colored red; (b) The spatial structure of dodecagonal tube made of edge-shared icosahedra; (c) Three edge-shared icosahedra forming the triple-join of dodecagonal ''rings''}
  \label{fig:tubes}
\end{figure}

As reported in Ref.~\cite{Dzugutov1993PRL}, the DQC structure of the system with Dzugutov potential is locally icosahedral. We have checked that structure of other systems studied are the same as obtained by Dzugutov. For example, in Fig.~\ref{fig:tubes}a  we show the typical fragment of DQC tiling for RSS; the particles which are the centers of icosahedra are colored red and one of such icosahedron is marked by interparticle bonds. Such icosahedron is the screen plain projection of spatial tube structure made of edge-shared icosahedra \cite{Dzugutov1993PRL} (see Fig.~\ref{fig:tubes}b). So the red particles in Fig.~\ref{fig:tubes}a represent the axes of these tubes. As seen from Fig.~\ref{fig:snaphsots} and Fig.~\ref{fig:tubes}a, there are two joining mechanisms of dodecagonal rings: triple and quadruple junctions. In Fig.~\ref{fig:tubes}c we show local structure of the triple one made of three face-shared icosahedra.

 Note that earlier we reported the decagonal QC formation for RSS as well as for other two-length scale systems \cite{Ryltsev2015SoftMatt}. The building block of such quasicrystals is the similar icosahedral tube as showed in Fig.~\ref{fig:tubes}b but made of face-shared icosahedra.

\subsection{The origin of universality}

\begin{figure}
  \centering
  \includegraphics[width=0.9\columnwidth]{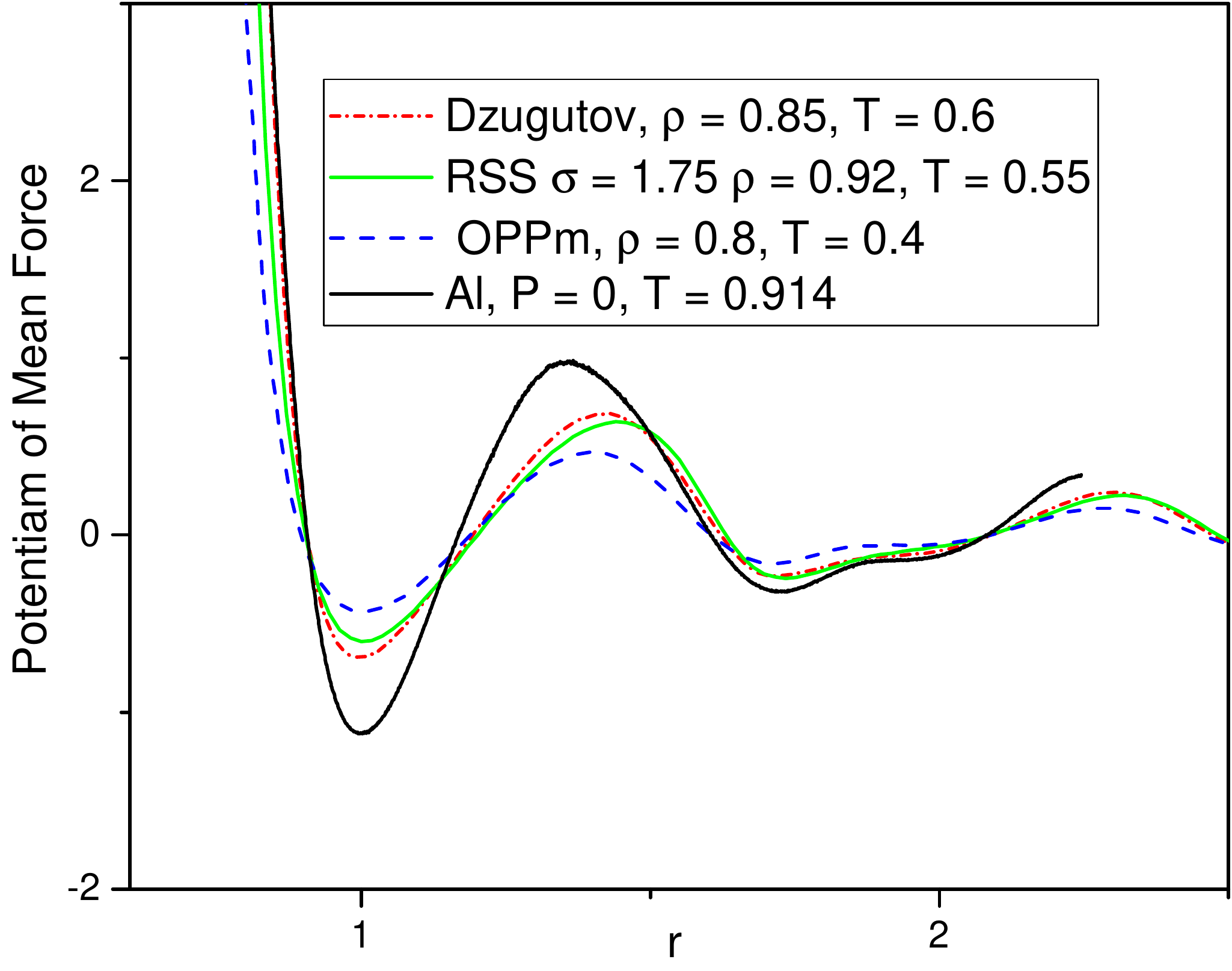}\\
 \caption{ The potentials of mean force $U_{\rm pmf}(r)=-kT\ln(g(r))$ for the systems under investigation. The radial distribution functions $g(r)$ used to calculate $U_{\rm pmf}$ are the same as presented in Fig.~\ref{fig:rdf_pot}b.}
  \label{fig:pmf}
\end{figure}

We have shown above that two-length-scale systems of different nature demonstrate the same DQC structure with similar values of effective parameters characterizing the fluid structure. It suggests common mechanism of fluid-DQC transition in the systems under consideration. Even though it is obvious that two-scale nature of the interparticle interaction plays an important role in QC formation, the origin of such universality is not completely clear. The very fact that effective parameters are extracted from radial distribution function $g(r)$ suggests that different systems with similar $g(r)$ in fluid phase form the same solid structures under cooling. This idea is supported by the fact that $g(r)$ determines pair potential of mean force (PMF) $U_{\rm pmf}(r)=-kT\ln(g(r))$ that is the function whose gradient gives the force between two particles averaged over the equilibrium distribution of over particles \cite{hansenMcDonald, chandler1988stat_mech}. It is natural to guess that similarity of such effective forces in the fluid phase leads to similarity of solid state structure. To support this idea, we show in Fig.~\ref{fig:pmf} PMFs for the systems under consideration. We see that $U_{\rm pmf}$ calculated at thermodynamic states near fluid-DQC transition are very close to each other. Note that PMF for aluminum differs noticeable from those for other systems studied. As the consequence the aluminum has the values of effective parameters which are slightly less than optimal ones (Fig.~\ref{fig:rdf_pot}) and demonstrates worse DQC structure (Fig.~\ref{fig:snaphsots}).

\section{Conclusions}

In summary we show by molecular dynamic simulations that two-length-scale systems of different nature, both metallic- and soft-matter-like, can form the same DQC phases. That suggests that mechanism of DQC formation is universal for both metallic and soft-matter systems and it is based on competition between interparticle scales. We propose the universal criterion for DQC formation based on the values of the two effective dimensionless parameters extracted from the radial distribution function of the system in the fluid state near the fluid-DQC transition. The parameters reflect the existence of two effective interparticle distances (bond lengthes) originated from two-length-scale nature of interaction potential. These are the ratio between effective bond lengthes, $\lambda$, and the fraction of short-bonded particles $\phi$. The parameter values favoring the dodecagonal ordering was estimated to be $\lambda \sim 1.7$, $\phi\sim 0.4$. The proposed method allows reducing the time spent for searching the parameters favoring certain solid structure for given system. Indeed, simulation of the fluid state, where we get the effective parameters, requires much  less computational expenses than direct simulation of a fluid-solid transition.

\section{Acknowledgments}
 This work was supported by Russian Science Foundation (grant No. 14-13-00676). We are grateful to Ural Branch of Russian Academy of Sciences, Joint Supercomputer Center of the Russian Academy of Sciences and Federal Center for collective usage at NRC ''Kurchatov Institute'' for the access to supercomputer resources.




\bibliography{our_bib_qc} 

\providecommand*{\mcitethebibliography}{\thebibliography}
\csname @ifundefined\endcsname{endmcitethebibliography}
{\let\endmcitethebibliography\endthebibliography}{}
\begin{mcitethebibliography}{49}
\providecommand*{\natexlab}[1]{#1}
\providecommand*{\mciteSetBstSublistMode}[1]{}
\providecommand*{\mciteSetBstMaxWidthForm}[2]{}
\providecommand*{\mciteBstWouldAddEndPuncttrue}
  {\def\EndOfBibitem{\unskip.}}
\providecommand*{\mciteBstWouldAddEndPunctfalse}
  {\let\EndOfBibitem\relax}
\providecommand*{\mciteSetBstMidEndSepPunct}[3]{}
\providecommand*{\mciteSetBstSublistLabelBeginEnd}[3]{}
\providecommand*{\EndOfBibitem}{}
\mciteSetBstSublistMode{f}
\mciteSetBstMaxWidthForm{subitem}
{(\emph{\alph{mcitesubitemcount}})}
\mciteSetBstSublistLabelBeginEnd{\mcitemaxwidthsubitemform\space}
{\relax}{\relax}

\bibitem[Stadnik(2012)]{stadnik2012QC}
Z.~M. Stadnik, \emph{Physical properties of quasicrystals}, Springer Science \&
  Business Media, 2012, vol. 126\relax
\mciteBstWouldAddEndPuncttrue
\mciteSetBstMidEndSepPunct{\mcitedefaultmidpunct}
{\mcitedefaultendpunct}{\mcitedefaultseppunct}\relax
\EndOfBibitem
\bibitem[Tsai(2008)]{Tsai2008SciTechAdvMat}
A.~P. Tsai, \emph{Sci. Technol. Adv. Mat.}, 2008, \textbf{9}, 013008\relax
\mciteBstWouldAddEndPuncttrue
\mciteSetBstMidEndSepPunct{\mcitedefaultmidpunct}
{\mcitedefaultendpunct}{\mcitedefaultseppunct}\relax
\EndOfBibitem
\bibitem[Zeng \emph{et~al.}(2004)Zeng, Ungar, Liu, Percec, Dulcey, and
  Hobbs]{Zeng2004Nature}
X.~Zeng, G.~Ungar, Y.~Liu, V.~Percec, A.~E. Dulcey and J.~K. Hobbs,
  \emph{Nature}, 2004, \textbf{428}, 157--160\relax
\mciteBstWouldAddEndPuncttrue
\mciteSetBstMidEndSepPunct{\mcitedefaultmidpunct}
{\mcitedefaultendpunct}{\mcitedefaultseppunct}\relax
\EndOfBibitem
\bibitem[Fischer \emph{et~al.}(2011)Fischer, Exner, Zielske, Perlich, Deloudi,
  Steurer, Lindner, and F\"{o}rster]{Fischer2011PNAS}
S.~Fischer, A.~Exner, K.~Zielske, J.~Perlich, S.~Deloudi, W.~Steurer,
  P.~Lindner and S.~F\"{o}rster, \emph{PNAS}, 2011, \textbf{108},
  1810--1814\relax
\mciteBstWouldAddEndPuncttrue
\mciteSetBstMidEndSepPunct{\mcitedefaultmidpunct}
{\mcitedefaultendpunct}{\mcitedefaultseppunct}\relax
\EndOfBibitem
\bibitem[Hayashida \emph{et~al.}(2007)Hayashida, Dotera, Takano, and
  Matsushita]{Hayashida2007PRL}
K.~Hayashida, T.~Dotera, A.~Takano and Y.~Matsushita, \emph{Phys. Rev. Lett.},
  2007, \textbf{98}, 195502\relax
\mciteBstWouldAddEndPuncttrue
\mciteSetBstMidEndSepPunct{\mcitedefaultmidpunct}
{\mcitedefaultendpunct}{\mcitedefaultseppunct}\relax
\EndOfBibitem
\bibitem[Talapin \emph{et~al.}(2009)Talapin, Shevchenko, Bodnarchuk, Ye, Chen,
  and Murray]{Talapin2009Nature}
D.~V. Talapin, E.~V. Shevchenko, M.~I. Bodnarchuk, X.~Ye, J.~Chen and C.~B.
  Murray, \emph{Nature}, 2009, \textbf{461}, 964--967\relax
\mciteBstWouldAddEndPuncttrue
\mciteSetBstMidEndSepPunct{\mcitedefaultmidpunct}
{\mcitedefaultendpunct}{\mcitedefaultseppunct}\relax
\EndOfBibitem
\bibitem[Zaidouny \emph{et~al.}(2014)Zaidouny, Bohlein, Roth, and
  Bechinger]{Zaidouny2014SoftMatt}
L.~Zaidouny, T.~Bohlein, J.~Roth and C.~Bechinger, \emph{Soft Matter}, 2014,
  \textbf{10}, 8705--8710\relax
\mciteBstWouldAddEndPuncttrue
\mciteSetBstMidEndSepPunct{\mcitedefaultmidpunct}
{\mcitedefaultendpunct}{\mcitedefaultseppunct}\relax
\EndOfBibitem
\bibitem[Ungar and Zeng(2005)]{Ungar2005SoftMatt}
G.~Ungar and X.~Zeng, \emph{Soft Matter}, 2005, \textbf{1}, 95--106\relax
\mciteBstWouldAddEndPuncttrue
\mciteSetBstMidEndSepPunct{\mcitedefaultmidpunct}
{\mcitedefaultendpunct}{\mcitedefaultseppunct}\relax
\EndOfBibitem
\bibitem[Denton and L\"owen(1998)]{Denton1998PRL}
A.~R. Denton and H.~L\"owen, \emph{Phys. Rev. Lett.}, 1998, \textbf{81},
  469--472\relax
\mciteBstWouldAddEndPuncttrue
\mciteSetBstMidEndSepPunct{\mcitedefaultmidpunct}
{\mcitedefaultendpunct}{\mcitedefaultseppunct}\relax
\EndOfBibitem
\bibitem[Denton and Hafner(1997)]{Denton1997PRB}
A.~R. Denton and J.~Hafner, \emph{Phys. Rev. B}, 1997, \textbf{56},
  2469--2482\relax
\mciteBstWouldAddEndPuncttrue
\mciteSetBstMidEndSepPunct{\mcitedefaultmidpunct}
{\mcitedefaultendpunct}{\mcitedefaultseppunct}\relax
\EndOfBibitem
\bibitem[Dzugutov(1993)]{Dzugutov1993PRL}
M.~Dzugutov, \emph{Phys. Rev. Lett.}, 1993, \textbf{70}, 2924--2927\relax
\mciteBstWouldAddEndPuncttrue
\mciteSetBstMidEndSepPunct{\mcitedefaultmidpunct}
{\mcitedefaultendpunct}{\mcitedefaultseppunct}\relax
\EndOfBibitem
\bibitem[Engel \emph{et~al.}(2015)Engel, Damasceno, Phillips, and
  Glotzer]{Englel2015Nature}
M.~Engel, P.~F. Damasceno, C.~L. Phillips and S.~C. Glotzer, \emph{Nat.
  Mater.}, 2015, \textbf{14}, 109--116\relax
\mciteBstWouldAddEndPuncttrue
\mciteSetBstMidEndSepPunct{\mcitedefaultmidpunct}
{\mcitedefaultendpunct}{\mcitedefaultseppunct}\relax
\EndOfBibitem
\bibitem[Ryltsev \emph{et~al.}(2015)Ryltsev, Klumov, and
  Chtchelkatchev]{Ryltsev2015SoftMatt}
R.~Ryltsev, B.~Klumov and N.~Chtchelkatchev, \emph{Soft Matter}, 2015,
  \textbf{11}, 6991--6998\relax
\mciteBstWouldAddEndPuncttrue
\mciteSetBstMidEndSepPunct{\mcitedefaultmidpunct}
{\mcitedefaultendpunct}{\mcitedefaultseppunct}\relax
\EndOfBibitem
\bibitem[Damasceno \emph{et~al.}(2017)Damasceno, Glotzer, and
  Engel]{Damasceno2017JPCM}
P.~Damasceno, S.~Glotzer and M.~Engel, \emph{J. Phys.: Condens. Matter},
  2017\relax
\mciteBstWouldAddEndPuncttrue
\mciteSetBstMidEndSepPunct{\mcitedefaultmidpunct}
{\mcitedefaultendpunct}{\mcitedefaultseppunct}\relax
\EndOfBibitem
\bibitem[Metere \emph{et~al.}(2016)Metere, Oleynikov, Dzugutov, and
  Lidin]{Metere2016SoftMatter}
A.~Metere, P.~Oleynikov, M.~Dzugutov and S.~Lidin, \emph{Soft Matter}, 2016,
  \textbf{12}, 8869--8875\relax
\mciteBstWouldAddEndPuncttrue
\mciteSetBstMidEndSepPunct{\mcitedefaultmidpunct}
{\mcitedefaultendpunct}{\mcitedefaultseppunct}\relax
\EndOfBibitem
\bibitem[Lee(1981)]{Lee1981Book}
J.~K. Lee, \emph{Interatomic Potentials and Crystalline Defects.}, The
  metallurgical society of aime, warrendale, pa technical report, 1981\relax
\mciteBstWouldAddEndPuncttrue
\mciteSetBstMidEndSepPunct{\mcitedefaultmidpunct}
{\mcitedefaultendpunct}{\mcitedefaultseppunct}\relax
\EndOfBibitem
\bibitem[Mitra(1978)]{Mitra1978JPhysC}
S.~Mitra, \emph{J. Phys. C}, 1978, \textbf{11}, 3551\relax
\mciteBstWouldAddEndPuncttrue
\mciteSetBstMidEndSepPunct{\mcitedefaultmidpunct}
{\mcitedefaultendpunct}{\mcitedefaultseppunct}\relax
\EndOfBibitem
\bibitem[Mihalkovi\ifmmode~\check{c}\else \v{c}\fi{} and
  Henley(2012)]{Mihalkovich2012PRB}
M.~Mihalkovi\ifmmode~\check{c}\else \v{c}\fi{} and C.~L. Henley, \emph{Phys.
  Rev. B}, 2012, \textbf{85}, 092102\relax
\mciteBstWouldAddEndPuncttrue
\mciteSetBstMidEndSepPunct{\mcitedefaultmidpunct}
{\mcitedefaultendpunct}{\mcitedefaultseppunct}\relax
\EndOfBibitem
\bibitem[Dubinin \emph{et~al.}(2014)Dubinin, Filippov, Yuryev, and
  Vatolin]{Dubinin2014JNonCrystSol}
N.~E. Dubinin, V.~V. Filippov, A.~A. Yuryev and N.~A. Vatolin, \emph{J.
  Non-Cryst. Solids}, 2014, \textbf{401}, 101--104\relax
\mciteBstWouldAddEndPuncttrue
\mciteSetBstMidEndSepPunct{\mcitedefaultmidpunct}
{\mcitedefaultendpunct}{\mcitedefaultseppunct}\relax
\EndOfBibitem
\bibitem[Dubinin \emph{et~al.}(2014)Dubinin, Vatolin, and
  Filippov]{Dubinin2014RusChemRev}
N.~E. Dubinin, N.~A. Vatolin and V.~V. Filippov, \emph{Russ. Chem. Rev.}, 2014,
  \textbf{83}, 987\relax
\mciteBstWouldAddEndPuncttrue
\mciteSetBstMidEndSepPunct{\mcitedefaultmidpunct}
{\mcitedefaultendpunct}{\mcitedefaultseppunct}\relax
\EndOfBibitem
\bibitem[Likos(2001)]{Likos2001PhysRep}
C.~N. Likos, \emph{Phys. Rep.}, 2001, \textbf{348}, 267--439\relax
\mciteBstWouldAddEndPuncttrue
\mciteSetBstMidEndSepPunct{\mcitedefaultmidpunct}
{\mcitedefaultendpunct}{\mcitedefaultseppunct}\relax
\EndOfBibitem
\bibitem[Watzlawek \emph{et~al.}(1999)Watzlawek, Likos, and
  L\"owen]{Watzlawek1999PRL}
M.~Watzlawek, C.~N. Likos and H.~L\"owen, \emph{Phys. Rev. Lett.}, 1999,
  \textbf{82}, 5289--5292\relax
\mciteBstWouldAddEndPuncttrue
\mciteSetBstMidEndSepPunct{\mcitedefaultmidpunct}
{\mcitedefaultendpunct}{\mcitedefaultseppunct}\relax
\EndOfBibitem
\bibitem[Likos \emph{et~al.}(2002)Likos, Hoffmann, L\"owen, and
  Louis]{Likos2002JPhysCondMatt}
C.~N. Likos, N.~Hoffmann, H.~L\"owen and A.~A. Louis, \emph{J. Phys.: Condens.
  Matter}, 2002, \textbf{14}, 7681\relax
\mciteBstWouldAddEndPuncttrue
\mciteSetBstMidEndSepPunct{\mcitedefaultmidpunct}
{\mcitedefaultendpunct}{\mcitedefaultseppunct}\relax
\EndOfBibitem
\bibitem[Prestipino \emph{et~al.}(2009)Prestipino, Saija, and
  Malescio]{Prestipino2009SoftMatt}
S.~Prestipino, F.~Saija and G.~Malescio, \emph{Soft Matter}, 2009, \textbf{5},
  2795--2803\relax
\mciteBstWouldAddEndPuncttrue
\mciteSetBstMidEndSepPunct{\mcitedefaultmidpunct}
{\mcitedefaultendpunct}{\mcitedefaultseppunct}\relax
\EndOfBibitem
\bibitem[Rechtsman \emph{et~al.}(2006)Rechtsman, Stillinger, and
  Torquato]{Rechtsman2006PRE}
M.~Rechtsman, F.~Stillinger and S.~Torquato, \emph{Phys. Rev. E}, 2006,
  \textbf{73}, 011406\relax
\mciteBstWouldAddEndPuncttrue
\mciteSetBstMidEndSepPunct{\mcitedefaultmidpunct}
{\mcitedefaultendpunct}{\mcitedefaultseppunct}\relax
\EndOfBibitem
\bibitem[Ryltsev \emph{et~al.}(2013)Ryltsev, Chtchelkatchev, and
  Ryzhov]{Ryltsev2013PRL}
R.~E. Ryltsev, N.~M. Chtchelkatchev and V.~N. Ryzhov, \emph{Phys. Rev. Lett.},
  2013, \textbf{110}, 025701\relax
\mciteBstWouldAddEndPuncttrue
\mciteSetBstMidEndSepPunct{\mcitedefaultmidpunct}
{\mcitedefaultendpunct}{\mcitedefaultseppunct}\relax
\EndOfBibitem
\bibitem[Mishin \emph{et~al.}(1999)Mishin, Farkas, Mehl, and
  Papaconstantopoulos]{Mishin1999PRB}
Y.~Mishin, D.~Farkas, M.~J. Mehl and D.~A. Papaconstantopoulos, \emph{Phys.
  Rev. B}, 1999, \textbf{59}, 3393--3407\relax
\mciteBstWouldAddEndPuncttrue
\mciteSetBstMidEndSepPunct{\mcitedefaultmidpunct}
{\mcitedefaultendpunct}{\mcitedefaultseppunct}\relax
\EndOfBibitem
\bibitem[Prokhoda and Ovrutsky(2014)]{Prokhoda2014arXiv}
A.~Prokhoda and A.~Ovrutsky, \emph{arXiv:1403.6668}, 2014\relax
\mciteBstWouldAddEndPuncttrue
\mciteSetBstMidEndSepPunct{\mcitedefaultmidpunct}
{\mcitedefaultendpunct}{\mcitedefaultseppunct}\relax
\EndOfBibitem
\bibitem[Dzugutov(1992)]{Dzugutov1992PRA}
M.~Dzugutov, \emph{Phys. Rev. A}, 1992, \textbf{46}, R2984--R2987\relax
\mciteBstWouldAddEndPuncttrue
\mciteSetBstMidEndSepPunct{\mcitedefaultmidpunct}
{\mcitedefaultendpunct}{\mcitedefaultseppunct}\relax
\EndOfBibitem
\bibitem[Keys and Glotzer(2007)]{Keys2007PRL}
A.~S. Keys and S.~C. Glotzer, \emph{Phys. Rev. Lett.}, 2007, \textbf{99},
  235503\relax
\mciteBstWouldAddEndPuncttrue
\mciteSetBstMidEndSepPunct{\mcitedefaultmidpunct}
{\mcitedefaultendpunct}{\mcitedefaultseppunct}\relax
\EndOfBibitem
\bibitem[Roth and Denton(2000)]{Denton2000PRE}
J.~Roth and A.~R. Denton, \emph{Phys. Rev. E}, 2000, \textbf{61},
  6845--6857\relax
\mciteBstWouldAddEndPuncttrue
\mciteSetBstMidEndSepPunct{\mcitedefaultmidpunct}
{\mcitedefaultendpunct}{\mcitedefaultseppunct}\relax
\EndOfBibitem
\bibitem[Young and Alder(1977)]{Young1977PRL}
D.~A. Young and B.~J. Alder, \emph{Phys. Rev. Lett.}, 1977, \textbf{38},
  1213--1216\relax
\mciteBstWouldAddEndPuncttrue
\mciteSetBstMidEndSepPunct{\mcitedefaultmidpunct}
{\mcitedefaultendpunct}{\mcitedefaultseppunct}\relax
\EndOfBibitem
\bibitem[Stishov(2002)]{Stishov2002PhilMag}
S.~M. Stishov, \emph{Philos. Mag. B}, 2002, \textbf{82}, 1287--1290\relax
\mciteBstWouldAddEndPuncttrue
\mciteSetBstMidEndSepPunct{\mcitedefaultmidpunct}
{\mcitedefaultendpunct}{\mcitedefaultseppunct}\relax
\EndOfBibitem
\bibitem[lam()]{lammps}
\url{http://lammps.sandia.gov/}\relax
\mciteBstWouldAddEndPuncttrue
\mciteSetBstMidEndSepPunct{\mcitedefaultmidpunct}
{\mcitedefaultendpunct}{\mcitedefaultseppunct}\relax
\EndOfBibitem
\bibitem[Plimpton(1995)]{Plimpton1995JCompPhys}
S.~Plimpton, \emph{J. Comput. Phys.}, 1995, \textbf{117}, 1 -- 19\relax
\mciteBstWouldAddEndPuncttrue
\mciteSetBstMidEndSepPunct{\mcitedefaultmidpunct}
{\mcitedefaultendpunct}{\mcitedefaultseppunct}\relax
\EndOfBibitem
\bibitem[Kuksin \emph{et~al.}(2005)Kuksin, Morozov, Norman, Stegailov, and
  Valuev]{Kuksin2005MolSim}
A.~Y. Kuksin, I.~V. Morozov, G.~E. Norman, V.~V. Stegailov and I.~A. Valuev,
  \emph{Mol. Simulat.}, 2005, \textbf{31}, 1005--1017\relax
\mciteBstWouldAddEndPuncttrue
\mciteSetBstMidEndSepPunct{\mcitedefaultmidpunct}
{\mcitedefaultendpunct}{\mcitedefaultseppunct}\relax
\EndOfBibitem
\bibitem[Norman and Stegailov(2001)]{Norman2001JETP}
G.~E. Norman and V.~V. Stegailov, \emph{JETP}, 2001, \textbf{92},
  879--886\relax
\mciteBstWouldAddEndPuncttrue
\mciteSetBstMidEndSepPunct{\mcitedefaultmidpunct}
{\mcitedefaultendpunct}{\mcitedefaultseppunct}\relax
\EndOfBibitem
\bibitem[Steinhardt \emph{et~al.}(1981)Steinhardt, Nelson, and
  Ronchetti]{Steinhardt1981PRL}
P.~J. Steinhardt, D.~R. Nelson and M.~Ronchetti, \emph{Phys. Rev. Lett.}, 1981,
  \textbf{47}, 1297--1300\relax
\mciteBstWouldAddEndPuncttrue
\mciteSetBstMidEndSepPunct{\mcitedefaultmidpunct}
{\mcitedefaultendpunct}{\mcitedefaultseppunct}\relax
\EndOfBibitem
\bibitem[Steinhardt \emph{et~al.}(1983)Steinhardt, Nelson, and
  Ronchetti]{Steinhardt1983PRB}
P.~J. Steinhardt, D.~R. Nelson and M.~Ronchetti, \emph{Phys. Rev. B}, 1983,
  \textbf{28}, 784--805\relax
\mciteBstWouldAddEndPuncttrue
\mciteSetBstMidEndSepPunct{\mcitedefaultmidpunct}
{\mcitedefaultendpunct}{\mcitedefaultseppunct}\relax
\EndOfBibitem
\bibitem[Hirata \emph{et~al.}(2013)Hirata, Kang, Fujita, Klumov, Matsue,
  Kotani, Yavari, and Chen]{Hirata2013Science}
A.~Hirata, L.~J. Kang, T.~Fujita, B.~Klumov, K.~Matsue, M.~Kotani, A.~R. Yavari
  and M.~W. Chen, \emph{Science}, 2013, \textbf{341}, 376--379\relax
\mciteBstWouldAddEndPuncttrue
\mciteSetBstMidEndSepPunct{\mcitedefaultmidpunct}
{\mcitedefaultendpunct}{\mcitedefaultseppunct}\relax
\EndOfBibitem
\bibitem[Butler and Hopkins(1970)]{Butler1980PhotocemPotobiol}
W.~L. Butler and D.~W. Hopkins, \emph{Photochem. Photobiol.}, 1970,
  \textbf{12}, 439--450\relax
\mciteBstWouldAddEndPuncttrue
\mciteSetBstMidEndSepPunct{\mcitedefaultmidpunct}
{\mcitedefaultendpunct}{\mcitedefaultseppunct}\relax
\EndOfBibitem
\bibitem[Aragao and Messaddeq(2008)]{Arag2008JBrazChemSoc}
B.~J. G.~d. Aragao and Y.~Messaddeq, \emph{{J. Braz. Chem. Soc.}}, 2008,
  \textbf{19}, 1582 -- 1594\relax
\mciteBstWouldAddEndPuncttrue
\mciteSetBstMidEndSepPunct{\mcitedefaultmidpunct}
{\mcitedefaultendpunct}{\mcitedefaultseppunct}\relax
\EndOfBibitem
\bibitem[Baker and Graves-Morris(1996)]{baker1996pade}
G.~A. Baker and P.~R. Graves-Morris, \emph{Pad{\'e} Approximants}, Cambridge
  University Press, 1996, vol.~59\relax
\mciteBstWouldAddEndPuncttrue
\mciteSetBstMidEndSepPunct{\mcitedefaultmidpunct}
{\mcitedefaultendpunct}{\mcitedefaultseppunct}\relax
\EndOfBibitem
\bibitem[Sch\"ott \emph{et~al.}(2016)Sch\"ott, van Loon, Locht, Katsnelson, and
  Di~Marco]{Schott2016PhysRevB}
J.~Sch\"ott, E.~G. C.~P. van Loon, I.~L.~M. Locht, M.~I. Katsnelson and
  I.~Di~Marco, \emph{Phys. Rev. B}, 2016, \textbf{94}, 245140\relax
\mciteBstWouldAddEndPuncttrue
\mciteSetBstMidEndSepPunct{\mcitedefaultmidpunct}
{\mcitedefaultendpunct}{\mcitedefaultseppunct}\relax
\EndOfBibitem
\bibitem[Chtchelkatchev and Ryltsev(2015)]{Chtchelkatchev2015JETPLett}
N.~Chtchelkatchev and R.~Ryltsev, \emph{JETP. Lett.}, 2015, \textbf{102},
  732--738\relax
\mciteBstWouldAddEndPuncttrue
\mciteSetBstMidEndSepPunct{\mcitedefaultmidpunct}
{\mcitedefaultendpunct}{\mcitedefaultseppunct}\relax
\EndOfBibitem
\bibitem[Chtchelkatchev \emph{et~al.}(2016)Chtchelkatchev, Klumov, Ryltsev,
  Khusnutdinoff, and Mokshin]{Chtchelkatchev2016JETPLett}
N.~M. Chtchelkatchev, B.~A. Klumov, R.~E. Ryltsev, R.~M. Khusnutdinoff and
  A.~V. Mokshin, \emph{JETP Letters}, 2016, \textbf{103}, 390--394\relax
\mciteBstWouldAddEndPuncttrue
\mciteSetBstMidEndSepPunct{\mcitedefaultmidpunct}
{\mcitedefaultendpunct}{\mcitedefaultseppunct}\relax
\EndOfBibitem
\bibitem[Khusnutdinoff \emph{et~al.}(2016)Khusnutdinoff, Mokshin, Klumov,
  Ryltsev, and Chtchelkatchev]{Khusnutdinoff2016JETP}
R.~M. Khusnutdinoff, A.~V. Mokshin, B.~A. Klumov, R.~E. Ryltsev and N.~M.
  Chtchelkatchev, \emph{JETP}, 2016, \textbf{123}, 265--276\relax
\mciteBstWouldAddEndPuncttrue
\mciteSetBstMidEndSepPunct{\mcitedefaultmidpunct}
{\mcitedefaultendpunct}{\mcitedefaultseppunct}\relax
\EndOfBibitem
\bibitem[Hansen and McDonald(2013)]{hansenMcDonald}
J.-P. Hansen and I.~R. McDonald, \emph{Theory of simple liquids}, Academic
  Press, Fourth Edition edn., 2013\relax
\mciteBstWouldAddEndPuncttrue
\mciteSetBstMidEndSepPunct{\mcitedefaultmidpunct}
{\mcitedefaultendpunct}{\mcitedefaultseppunct}\relax
\EndOfBibitem
\bibitem[Chandler and Percus(1988)]{chandler1988stat_mech}
D.~Chandler and J.~K. Percus, \emph{Introduction to modern statistical
  mechanics}, 1988\relax
\mciteBstWouldAddEndPuncttrue
\mciteSetBstMidEndSepPunct{\mcitedefaultmidpunct}
{\mcitedefaultendpunct}{\mcitedefaultseppunct}\relax
\EndOfBibitem
\end{mcitethebibliography}
\bibliographystyle{rsc} 

\end{document}